# Spectrum Occupancy Measurement: An Autocorrelation based Scanning Technique using USRP

Sriram Subramaniam, Hector Reyes and Naima Kaabouch
Electrical Engineering, University of North Dakota
Grand Forks, North Dakota, United States

*Abstract*—This paper presents a technique for scanning and evaluating the radio spectrum use. This technique determines the average occupancy of a channel over a specific duration. The technique was implemented using Software Defined Radio units and GNU Radio software. The survey was conducted in Grand Forks, North Dakota, over a frequency range of 824 MHz to 5.8 GHz. The results of this technique were compared to those of two existing techniques, energy detection and autocorrelation, that were also implemented. The results show that the proposed technique is more efficient at scanning the radio spectrum than the other two techniques.

## I. Introduction

With the increase of portable device utilization and ever-growing demand for greater wireless data transmission rates, an increasing demand for spectrum channels has been observed over the last decade. Conventionally, licensed spectrum channels are assigned for comparatively long time spans to license holders who may not continuously use them, creating an under-utilized spectrum. The inefficient use of spectrum resources has motivated researchers to look for advanced, innovative technologies that enable more efficient spectrum resource use [1].

Spectrum surveys have been conducted worldwide at different locations, covering both wide frequency ranges and specific licensed bands. Specific surveys have been conducted in the USA [2-3], New Zealand [4], Singapore [5], Ireland [6], Germany [7], and Spain [8]**.** In the USA, the spectral occupancy was found to be higher in coastal cities because of the presence of naval radars. A spectrum occupancy measurement was done in 2005 in the city of Chicago, Illinois, over a two-day period over the range of 30-3000 MHz using the energy detection method with fixed-threshold [9]. The authors observed a maximum occupancy of 70.9% between 54 to 88 MHz and virtually no occupancy between 1240 and 1850 MHz. In [10], the authors performed a three-year survey on frequencies ranging from 30 MHz-6 GHz. They also used the energy detection technique with a fixed-threshold.

Similarly, in [11], a measurement survey was conducted in Hull, UK, using energy detection technique for spectrum sensing. The results of this survey show a high occupancy in the GSM900 and GSM1800 bands due to broadcasting downlinks and less than 10% utilization for frequencies above 1 GHz. In [7], spectral occupancy measurements were performed both indoors and outdoors. This study showed that outdoor measurements had higher occupancy than indoor, due to less signal attenuation as the transmitters had direct line-of-sight with the receiver antennas.

In all of the above measurement surveys, the technique used for spectrum sensing was energy detection with fixed threshold. However, there are several limitations associated with these surveys. First, the noise is random; hence the threshold should be dynamic. Second, studies have shown that energy detection has a high rate of false alarms with improper setting of the threshold values [12].

In this paper, we present a technique for scanning and evaluating radio spectrum use. This technique determines the occupancy of a channel instantaneously or over an extended duration of time. The technique was implemented using Software Defined Radio units and GNU Radio software. The results of this technique were compared to those obtained using two other techniques, energy detection and autocorrelation.

## II. Methodology

We implemented three scanning techniques: energy detection [12], autocorrelation function at lag 1 [13], and correlation distance. All these techniques were processed sequentially using Universal Software Radio Peripheral (USRP) units and GNU Radio software along with computers used for storage. For each channel, the software (written in Python) scanned the spectrum to determine the presence or absence of a signal using the three techniques. Fig. 1 shows the experimental setup with the steps of the algorithm. The measurements were performed during several days over several weeks and months.

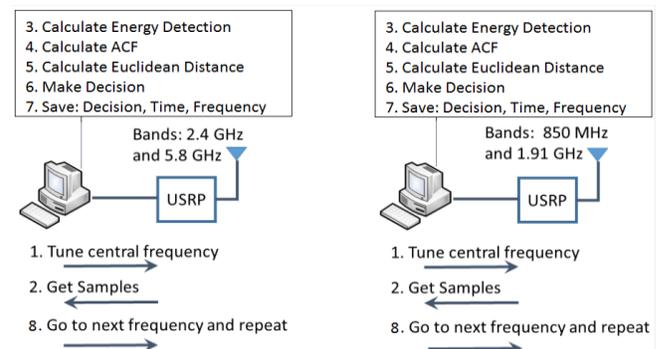

Fig. 1. Experimental Setup of Spectrum Sensing.

Table 1 shows the list of scanned frequency bands, their frequency ranges, their frequency steps, and the number of channels in each band. The other bands not listed in this table were identified as not significantly used.

TABLE I.   LIST OF SCANNED BANDS AND CHANNELS

| Band | Start-Stop Frequency (MHz) | Channel Spacing (MHz) | Number of channels |
|---|---|---|---|
| GSM-850 (U/L) | 824 – 849 | 3, 2 | 11 |
| GSM-850 (D/L) | 869 – 894 | 3, 2 | 11 |
| GSM-1900 (U/L) | 1850 – 1910 | 3, 2 | 25 |
| GSM-1900 (D/L) | 1930 – 1990 | 3, 2 | 25 |
| 2.4 GHz | 2402 – 2497 | 5 | 20 |
| 5.8 GHz | 5725 – 5875 | 5 | 31 |

During the scanning, results were stored in memory for further processing. After the scanning, for each channel, we calculated the average occupancy, $O_{average}$, for each of the three techniques over a specific duration that depends on the time of the day. This occupancy is defined as

$$O_{average} = \frac{N_{Detected}}{N_{Total}} \quad (1)$$

Where $N_{Detected}$ represents the number of detected signals over a duration $T$ and $N_{total}$ represents the total number of scans over the same duration. A brief description of each scanning technique is given below.

### A. Energy Detection

In its simplest form, energy detection computes the energy of the received signal $y(n)$ as a decision statistic $T_{ED}$ and then compares $T_{ED}$ with a predetermined fixed threshold $\lambda_{ED}$. The decision statistic can be expressed as

$$T_{ED} = \frac{1}{N}\sum_{n=1}^{N} |y(n)|^2 \quad (2)$$

Where $T_{ED}$ is the decision statistic, $y(n)$ is the sampled received signal, and $N$ is the total number of samples in a detection cycle. The decision statistic $T_{ED}$ can be calculated from the squared magnitude of the FFT averaged over $N$ samples.

The decision statistic $T_{ED}$ is computed for each sensing cycle of $N$ samples and is compared to the threshold $\lambda_{ED}$ to get the sensing result according to the following equation:

$T_{ED} < \lambda_{ED}$   Signal absent   (3)

$T_{ED} > \lambda_{ED}$   Signal present   (4)

### B. Autocorrelation function at lag 1 (ACF(1))

In this method, the autocorrelation of the samples at lag l is defined as

$$ACF(l) = \sum_{m=0}^{N_s-1} x(m)x^*(m-l) \quad (5)$$

Where $N_s$ is the number of samples, $l$ is the time lag to produce the time-shifted version of the received sample, and $x(m)$ and the symbol * represent the complex conjugate operation. If two successive values of an autocorrelation function of a signal are close to each other, then the signal is more correlated; if the values significantly differ from each other, then the signal is least correlated or uncorrelated.

### C. Correlation Distance

An alternative approach to dealing with the energy of the received signal samples is to exploit the inherent properties that exist in signals which distinguish them from noise. The autocorrelation function (ACF) is one operation that exploits such features. Since the additive white Gaussian noise is random, its ACF is highly uncorrelated. However, the ACF of a signal is correlated and the degree of the correlation defines the strength of the signal; the higher the degree of correlation, the greater the signal strength. In this proposed approach, we define a reference vector $ACF_{ref}$ which consists of auto correlated values of a signal, which are strong enough to have certainty about its presence. Another vector $ACF_{in}$ consists of auto correlated values of $N_s$ samples of the received signal. The correlation distance $D_{Correlation}$ is computed as the distance between the two vectors $ACF_{ref}$ and $ACF_{in}$ and is expressed as

$$D_{Correlation} = \sqrt{\sum (ACF_{ref} - ACF_{in})^2} \quad (6)$$

The $D_{Correlation}$ is the metric compared with a threshold $\gamma$ to decide about the presence of the signal. Repeated experiments yielded a threshold value $\gamma$ between 0 and 1. Any value of the $D_{Correlation}$ below $\gamma$ was a detected signal denoted with the binary value 1, and any value above $\gamma$ was an undetected signal with the binary value 0.

### III.   RESULTS

Examples of results are shown in Figs. 2 through 9. Each figure illustrate the occupancy of a particular channel in a selected band using the three aforementioned techniques (plots a, b, and c). Sub-plot a illustrates the channel occupancy measurement performed using energy detection technique with fixed threshold; sub-plot b illustrates the occupancy measurements performed using the ACF at lag 1 technique; and sub-plot c illustrates the occupancy measurements performed using the correlation distance method.

Figs. 2, 3, and 4 illustrate the occupancies of channel 1 (2.412 GHz), channel 6 (2.437 GHz), and channel 11 (2.462 GHz) of the 2.4 GHz band. Fig. 2a shows the occupancy using the energy detection method. This figure shows that this channel is fully occupied (100%) at all times of the day, over the entire week. This high level of occupancy is attributed to the small value of the static threshold and the high false alarm rate of the energy detection method. Fig. 2b illustrates the occupancy of the same channel (2.412 GHz) using ACF at lag 1, showing varying occupancy at different times of the day, which is an expected behavior. This method results in higher

occupancy values than expected, as it relies on only the first lag of the autocorrelation. Fig. 2c shows the occupancy of the channel based on correlation distance. The occupancy values appear to be more realistic and expected as compared to those of the above two approaches, with high occupancy values during the peak usage hours of 12pm – 4pm (20% - 40%). This method owes its accuracy to the signal detection reliance on all the lags/points of the autocorrelation.

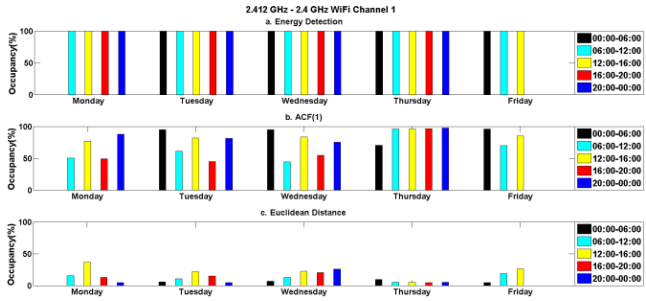

Fig. 2. Average occupancy of channel 1 (2.412 GHz) of 2.4 GHz Wi-Fi band.

Fig. 3 illustrates the occupancy levels of channel 6 (2.437 GHz) of the 2.4 GHz band. The results of the occupancy levels measured with the energy detection method (fig. 3a) is similar to that of channel 1 (2.412 GHz) as shown in fig. 2a. With respect to the ACF at lag 1 (fig. 3b), we see that there is a slight variation in the occupancy levels as compared to the occupancy levels of channel 1. Comparing figs. 2c and 3c, a distinguishing result can be observed with the occupancy results of the correlation distance method of measurement. It is evident that the overall usage of channel 6 is less than that of channel 1.

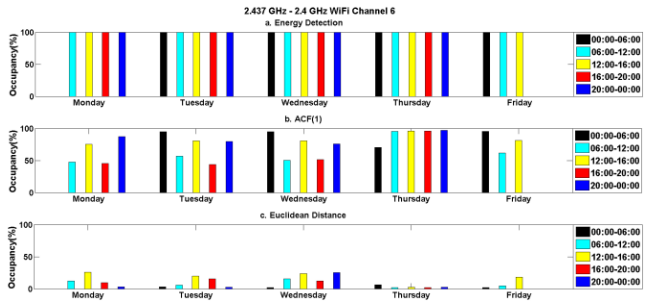

Fig. 3. Average occupancy of channel 6 (2.437 GHz) of 2.4 GHz Wi-Fi band.

The next highly occupied channel in the 2.4 GHz band is channel 11 (2.462 GHz). The results of the occupancy measurements with respect to the energy detection (fig. 4a) and ACF at lag 1 (fig. 4b) are similar to those of the previously mentioned channels 1 and 6, with slight variations in occupancy levels measured using ACF at lag 1 method. The distinguishing result is noticeable in the measurement performed using the correlation distance method as shown in fig. 4c, wherein a higher occupancy is noticed overall when compared to that of channel 6 (2.437 GHz). From the analysis of the results of the 2.4 GHz band, it is evident that channels 1, 6, and 11 are the most occupied channels of the 2.4 GHz band; channel 1 is the most occupied, followed by channel 11 and then by channel 6.

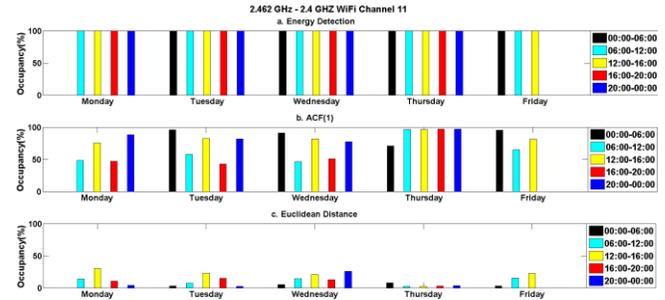

Fig. 4. Average occupancy of channel 11 (2.462 GHz) of 2.4 GHz Wi-Fi band.

In the next set of occupancy results, we will analyze the 5.8 GHz band and then the cellular bands such as GSM850 and GSM1900. Fig. 5 illustrates the results of the occupancy measurements of channel 153 (5.765 GHz) of the 5.8 GHz band. As can be seen in fig. 5a, the results corresponding to energy detection show constant 100% occupancy while those corresponding to the autocorrelation at lag 1, shown in fig. 5b, are lower and vary with time. On the other hand, the measurement of the occupancy values corresponding to the correlation distance, as shown in fig. 5c, are more realistic and expected as compared to those of the above two approaches.

Owing to the public holiday on New Year's Day (January 1, 2015), the correlation distance method demonstrates low-to-no activity on that day, hence demonstrating a more precise and accurate technique of signal detection. We infer that occupancy is highest in the 12pm – 4pm interval over the week and is relatively low when compared to the 2.4 GHz band, which is expected behavior.

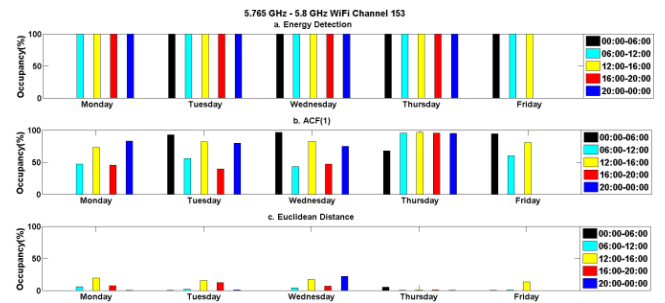

Fig. 5. Average occupancy of channel 153 (5.765 GHz) of 5.8 GHz Wi-Fi band.

Occupancies of the GSM850 and GSM1900 bands were also measured; their occupancy results are illustrated in figs. 6, 7, 8, and 9. Figs. 6 and 7 show the occupancies of the uplink (837 MHz) and downlink (882 MHz) channels, 192 of the GSM850 band. Both these channels (uplink and downlink) demonstrate 100% occupancy for all the three spectrum sensing techniques.

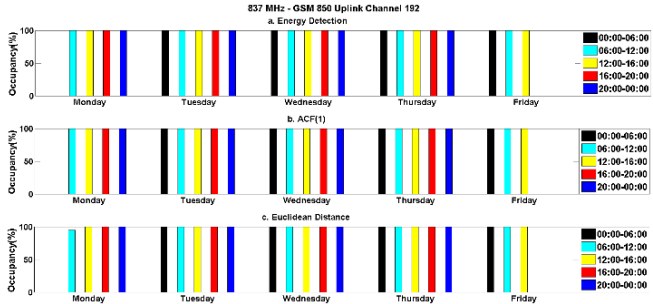
Fig. 6. Average occupancy of channel 192 (837 MHz) of GSM-850 band.

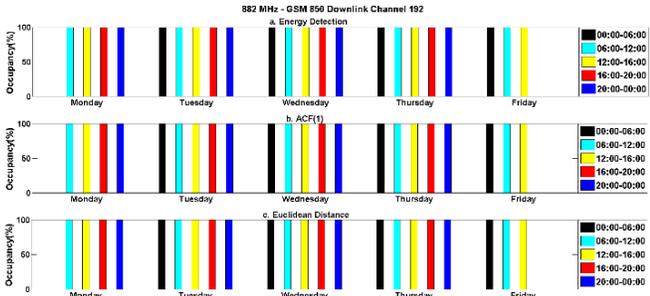
Fig. 7. Average occupancy of channel 192 (882 MHz) of GSM-850 band.

Figs. 8 and 9 depict the occupancies of the uplink and downlink channel 661 of the GSM-1900 band. Due to broadcasting downlink, the downlink (1960 MHz) channel 661 of the GSM1900 band demonstrates a 100% usage for all the three techniques, while the uplink (1880 MHz) channel 661 of the GSM1900 demonstrates a low occupancy with the correlation distance technique and high occupancies with ACF(1) ($\geq$60%) and Energy Detection (100%).

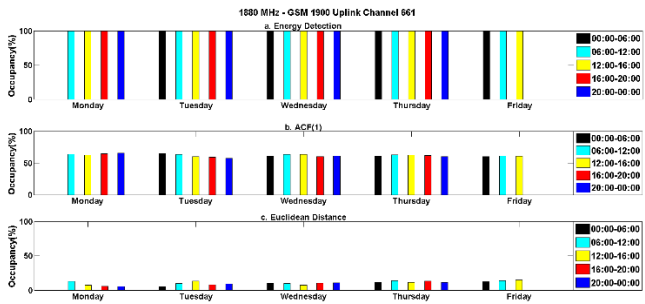
Fig. 8. Average occupancy of channel 661 (1880 MHz) of GSM-1900 band.

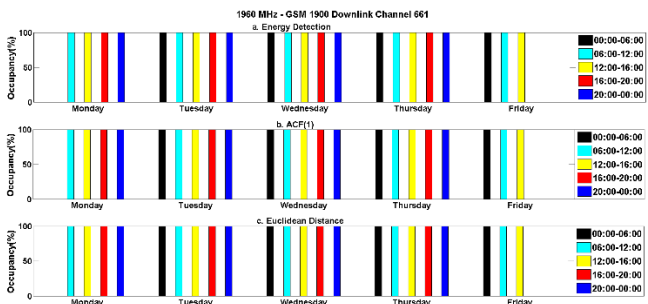
Fig. 9. Average occupancy of channel 661 (1960 MHz) of GSM-1900 band.

## IV. CONCLUSION

In this paper, we described a scanning technique and compared its performance to two other techniques, autocorrelation at lag 1 and energy detection. The experiments were performed on the radio spectrum over a frequency range of 824 MHz to 5.8 GHz in Grand Forks, North Dakota. As expected, the spectrum occupancy of any channel was found to be less than 20% in certain bands of the radio spectrum. The results also show that the occupancy changes depending on the time, day, and channel. Moreover, these results show that the proposed technique is more efficient at detecting signals than the other techniques.

ACKNOWLEDGMENT

The authors acknowledge the support of NSF, grant # 1443861, and EPSCoR/NSF, grant # EPS-0184442.

REFERENCES

[1] N. Kaabouch and WC Hu, Software-Defined and Cognitive Radio Technologies for Dynamic Spectrum Management, IGI Global, Volumes I and II, October 2014.
[2] F. H. Sanders, "Broadband spectrum surveys in Denver, CO, San Diego, CA, and Los Angeles, CA: methodology, analysis, and comparative results," in IEEE International Symposium on Electromagnetic Compatibility, 1998, pp. 988-993 vol.2.
[3] J. Do, D. M. Akos, and P. K. Enge, "L and S bands spectrum survey in the San Francisco bay area," in Position Location and Navigation Symposium, 2004, pp. 566-572.
[4] R. I. Chiang, G. B. Rowe, and K. W. Sowerby, "A quantitative analysis of spectral occupancy measurements for cognitive radio," in IEEE 65th Vehicular Technology Conference, 2007, pp. 3016-3020.
[5] M. H. Islam, C. L. Koh, S. W. Oh, X. Qing, Y. Y. Lai, C. Wang, et al., "Spectrum survey in Singapore: Occupancy measurements and analyses," in 3rd International Conference on Cognitive Radio Oriented Wireless Networks and Communications, 2008, pp. 1-7.
[6] Shared Spectrum Company, "Spectrum occupancy measurements," Shared Spectrum Company Reports (Jan 2004 - Aug 2005). Available at: http://www.sharedspectrum.com/measurements/.
[7] M. Wellens, J. Wu, and P. Mahonen, "Evaluation of spectrum occupancy in indoor and outdoor scenario in the context of cognitive radio," in 2nd International Conference on Cognitive Radio Oriented Wireless Networks and Communications, 2007, pp. 420-427.
[8] M. Lopez-Benitez, A. Umbert, and F. Casadevall, "Evaluation of spectrum occupancy in Spain for cognitive radio applications," in IEEE 69th Vehicular Technology Conference, 2009, pp. 1-5.
[9] M. A. McHenry, P. A. Tenhula, D. McCloskey, D. A. Roberson, and C. S. Hood, "Chicago spectrum occupancy measurements & analysis and a long-term studies proposal," presented at the Proceedings of the first international workshop on Technology and policy for accessing spectrum, Boston, Massachusetts, 2006.
[10] T. M. Taher, R. B. Bacchus, K. J. Zdunek, and D. A. Roberson, "Long-term spectral occupancy findings in Chicago," in IEEE Symposium on New Frontiers in Dynamic Spectrum Access Networks (DySPAN), 2011, pp. 100-107.
[11] M. Mehdawi, N. Riley, K. Paulson, A. Fanan, and M. Ammar, "Spectrum Occupancy Survey In HULL-UK For Cognitive Radio Applications: Measurement & Analysis," International Journal of Sceintific & Technology research, vol. 2, 2013.
[12] S. Saleem and K. Shahzad, "Performance evaluation of energy detection based spectrum sensing technique for wireless channel," International Journal of Multidisciplinary Sciences and Engineering, 2012, vol. 3, pp. 31-34.

[13] R. K. Sharma and J. W. Wallace, "Improved autocorrelation-based sensing using correlation distribution information," in International ITG Workshop on Smart Antennas (WSA), 2010, pp. 335-341.